\documentclass[a4paper, amsfonts, amssymb, amsmath, reprint, showkeys, nofootinbib, twoside]{revtex4-1}
\usepackage[english]{babel}
\usepackage[utf8]{inputenc}
\usepackage[colorinlistoftodos, color=green!40, prependcaption]{todonotes}
\usepackage{mathtools}
\usepackage{amsmath}
\usepackage{amssymb}

\usepackage{amsthm}
\usepackage{mathtools}
\usepackage{physics}
\usepackage{xcolor}
\usepackage{graphicx}
\usepackage[left=23mm,right=13mm,top=35mm,columnsep=15pt]{geometry} 
\usepackage{adjustbox}
\usepackage{placeins}
\usepackage[T1]{fontenc}
\usepackage{lipsum}
\usepackage{csquotes}
\usepackage{tikz}
\usetikzlibrary{quantikz}
\usepackage{adjustbox}

\usepackage[pdftex, pdftitle={Article}, pdfauthor={Author}]{hyperref} 
\begin{document}

\title{Bell State Analysis Provides an Optimal Basis Saturating the Quantum Cramer-Rao in Rotation Sensing}

\author{Zhuoran Bao}
    \email[Correspondence email address: ]{zhuoran.bao@mail.utoronto.ca}
    \affiliation{Dept. of Physics, University of Toronto, Toronto, M5S 1A7, Ontario, Canada}
\author{Daniel F. V. James}
    \email[Correspondence email address: ]{
    dfvj@physics.utoronto.ca}
    \affiliation{Dept. of Physics, University of Toronto, Toronto, M5S 1A7, Ontario, Canada}

\date{\today}

\begin{abstract}
The second-order anti-coherent state of light is known to saturate the Cramer-Rao Bound (QCRB) for rotation sensing around an arbitrary axis. However, due to the complexity of the state and the inefficiency of state tomography, parameter extraction remains an open problem. In this manuscript, we approach the problem of parameter extraction using pair-wise Bell state analysis with an additional path degree of freedom. Due to the transformation property of rotation, only the symmetric Bell states will show up in projection in the final state. We exploit this advantage to develop a scheme for extracting the rotation angle for N=4 and N=6 second-order anti-coherent states.
\end{abstract}
\keywords{Rotation Sensing, Second-order Anticoherent Polarization State, Cramer-Rao Bound, Bell State Analysis}

\maketitle

\section{Introduction}
Polarimetry is the study of polarization changes in an electromagnetic field after interaction and falls within a broader class of rotation-sensing problems \cite{Goldberg}. It has applications from biology to materials science. Its most prominent application lies in its ability to drastically improve imaging quality, allowing one to determine the chiral properties of chemical compounds, the surface reflectance of materials, and the imaging of living tissues \cite{Jessica, Sattar, Xiaodong, Rama}. Therefore, precise and efficient techniques for determining polarization rotations are in high demand. 

The simplest problem in rotation sensing is to estimate the rotation angle around a known axis. Polarization-entangled probe states, known as the polarization \textit{ Greenberger-Horne-Zeilinger} (GHZ) states \cite{Greenberger, Thomas}, analogous to NOON states for qubits, can theoretically achieve the Heisenberg scaling limit. In the small-rotation regime, the standard deviation of the measured phase scales as \(N^{-1}\) \cite{Giovannetti, Jaspreet, Jasminder}. To extract the rotation angle efficiently, a parity measurement must be performed at the end \cite{Higgins, Gao, Benjamin}. A classical alternative, which takes a longer time but yields the same amount of enhancement, exists: by allowing the light to repeatedly pass through the sample to accumulate the phase to achieve the enhancement \cite{Higgins}.

Determining the rotation around an unknown axis is a more interesting and challenging problem. Extensive research over the past decade has focused on finding quantum states that saturate the Quantum Cramer-Rao Bound (QCRB) \cite{Goldberg2, Gunnar, Björk, Martin}. It has been shown that the family of second-order unpolarized states, also known as second-order anti-coherence states, can saturate the QCRB if an optimal-basis analysis is performed. More precisely, using an N-qubit second-order anti-coherence state to measure the rotation parameters, one can achieve a precision that scales as \(\{N(N+2)\}^{-1/2}\approx N^{-1}\). Some effort has been put into generating these states \cite{Ferretti}; however, due to the lack of an optimal measurement procedure and the difficulty of generating anti-coherence states with high fidelity, no experiments using the optimal state for rotation sensing have been conducted yet.

In this paper, we propose a procedure that enables optimal basis measurement for the anti-coherence states using the path degree of freedom and linear optical components. We argue that since polarization rotation cannot alter the symmetry of the state, and the second-order anti-coherent states are symmetrical, the resultant state from the rotation can be decomposed into symmetrical Bell states. By adding the path degree of freedom to the existing polarization degree of freedom, Bell state analysis with linear optical components is sufficient for performing an optimal basis measurement in rotation sensing. 

The manuscript is organized into three parts. In Section II, we provide an overview of the notation we use, a brief review of Fisher information, and the QCRB. Then, in Section III, we examine the equivalence of optimal basis and the pair-wise Bell state analysis for saturating the QCRB and derive the basis explicitly for a four-qubit tetrahedron state and a six-qubit balanced state, both of which are second-order anti-coherence states. And finally, in Section IV, we examine the parameter-extraction procedure and use the multinomial distribution's variance to recover the Fisher information.

\section{The Quantum Fisher Information and the Optimal State for Rotation Sensing}
To start, let us define the notation used in the paper. In polarimetry, an N-photon polarization state refers to an N-qubit state obeying bosonic statistics. We can denote the N-photon polarization state as \(\vert\{ m, N-m\}\rangle\), where \(m\) and \(N-m\) are the number of horizontally and vertically polarized photons, respectively. We add the curly bracket to distinguish these two-mode number states from the angular momentum eigenstates introduced below. Such a state can be written as raising operators acting on the vacuum:
\begin{equation}
    \vert \{m,N-m\}\rangle=\frac{(\hat{a}^\dagger_H)^m(\hat{a}^\dagger_V)^{N-m}}{\sqrt{m!(N-m)!}}\vert vac\rangle.
\end{equation}
We can choose to write out the bosonic statistics as a permutation over arranging polarization qubits. Thus, it is equally valid to write the bosonic photon number state into a qubit state:
\begin{equation}
\begin{split}
    &\vert \{m, N-m\}\rangle \\
    &= \binom{N}{m}^{-\frac{1}{2}}\sum_{per}\hat{P} \left(\vert H\rangle^{\otimes m}\otimes\vert V\rangle^{\otimes (N-m)}\right).
\end{split}
\end{equation}
Here, \(\hat{P}\) is the permutation operator rearranging the photons, and the sum is over all possible permutations of arranging the horizontal and vertical photon states. This ability to convert the photon number state into qubit states will be important in making measurements in the optimal basis, which we will discuss in the next few sections.

Another representation that is very useful in our analysis is that these photon number states are analogous to the angular momentum states, where the total angular momentum is given by \(J=N/2\) with z-directional angular momentum \(m_j=(2m-N)/2\):
\begin{equation}
    \vert\{m,N-m\} \rangle\equiv\vert J,m_j\rangle=\left\vert \frac{N}{2},\frac{2m-N}{2}\right\rangle.
\end{equation}
Details of this analogy can be found in the literature \cite{Goldberg2}. We use the angular-momentum description for its simplicity; however, we will return to the qubit description when discussing the measurement process. As a cautious note, throughout this paper, we are only dealing with the "maximum spin" symmetric subspace of the entire N-qubit Hilbert space. For example, an \(N=4\) qubit space has the full dimension \(2^4=16\), thus has 16 basis states. When these bases are cast into the angular momentum basis representation, the full basis states are split into subspaces labelled by the total angular momentum \(J=0,\ 1,\ 2\). The polarization photon number states are only cast to the "maximum angular momentum" symmetric subspace \(J=2\). Explicitly, we are concerned with only the qubit state that can be written as a linear combination of the following basis states:
\begin{widetext}
    \begin{equation}
\begin{split}
    \vert \{4,0\}\rangle&\equiv\vert2,2\rangle\equiv\vert HHHH\rangle,\\
    \vert \{3,1\}\rangle&\equiv\vert2,1\rangle\equiv\frac{1}{2}\left(\vert HHHV\rangle+\vert HHVH\rangle+\vert HVHH\rangle+\vert VHHH\rangle\right),\\
    \vert \{2,2\}\rangle&\equiv\vert 2,0\rangle\equiv\frac{1}{\sqrt{6}}(\vert HHVV\rangle+\vert HVHV\rangle+\vert HVVH\rangle+\vert VHHV\rangle+\vert VHVH\rangle+\vert VVHH\rangle),\\
    \vert \{1,3\}\rangle&\equiv\vert 2,-1\rangle\equiv\frac{1}{2}(\vert VVVH\rangle+\vert VVHV\rangle+\vert VHVV\rangle+\vert HVVV\rangle),\\
    \vert\{0,4\}\rangle&\equiv\vert2,-2\rangle\equiv\vert VVVV\rangle.\\
\end{split}
\end{equation}
\end{widetext}

A second notation issue we wish to clarify concerns the unitary operator used to describe a polarization rotation. Suppose we have a pure N-photon polarization state and wish to perform a polarization rotation on it. In the N-qubit picture, let \(\hat{U}\) be an arbitrary unitary operation applied to a single qubit parameterized as the following:
\begin{equation}
    \hat{U}(\boldsymbol{\theta})=e^{-i\theta_1\boldsymbol{u}(\theta_2,\theta_3)\boldsymbol{\cdot\hat\sigma}}.
\end{equation}
Where \(\boldsymbol{\hat{\sigma}}\) is the Pauli matrix, \(\boldsymbol{u}(\theta_2,\theta_3)\) is the unit vector along the axis of rotation, \(\theta_1\) is the rotation angle. Explicitly, \(\boldsymbol{u}\) is parameterized as:
\begin{equation}\label{u}
    \boldsymbol{u}=\begin{pmatrix}
        \sin{\theta_2}\cos{\theta_3}\\
        \sin{\theta_2}\sin{\theta_3}\\
        \cos{\theta_2}
    \end{pmatrix}.
\end{equation}
A polarization rotation happens when an operation \(\hat{\mathcal{U}}=\hat{U}^{\otimes N}\) is applied to the N-qubit pure state. Following from the angular momentum analogy, such a polarization rotation is represented by:
\begin{equation}
    \hat{\mathcal{U}}(\boldsymbol{\theta})=e^{i\theta_1\boldsymbol{u}(\theta_2,\theta_3)\boldsymbol{\cdot\boldsymbol{\hat{J}}}}.
\end{equation}
Here, \(\boldsymbol{\hat{J}}\) is the angular momentum operator, which can be written as a sum of all permutations of Pauli matrices acting on a single qubit tensor with identity at other locations; explicitly, this is,
\begin{equation}
\begin{split}
    \hat{J}_j =& \sum_{per\ k} \hat{\sigma}_j^{(k)}\otimes\hat{I}^{\otimes N-1}
    =\hat{\sigma}_j\otimes\hat{I}^{\otimes N-1}\\&+\hat{I}\otimes\hat{\sigma}_j\otimes\hat{I}^{\otimes N-2}+...+\hat{I}^{\otimes N-1}\otimes\hat{\sigma}_j.
\end{split}
\end{equation}
Again, the sum is over all possible permutations of applying the Pauli \(\hat{\sigma}_j\) on the kth qubit while leaving the others unchanged. Then, \(\hat{\mathcal{U}}(\boldsymbol{\theta})\) is described by three independent parameters, \(\theta_1\) describing the rotation angle, and \(\theta_2,\ \theta_3\) denote the rotation axis \(\boldsymbol{u}(\theta_2,\theta_3)\). 

Let \(\hat{G}_1\) be the generators corresponding to the parameters \(\theta_1\) which features the rotation angle:
\begin{equation}\label{Gk}
    \hat{G}_1 = i\frac{\partial\hat{\mathcal{U}}}{\partial\theta_1}\hat{\mathcal{U}}^\dagger.
\end{equation}
The generators \(\hat{G}_2\) and \(\hat{G}_3\) corresponding to \(\theta_2\) and \(\theta_3\) can also be computed; however, they are much more difficult to compute due to the commutation relations between the Pauli operators. The details are presented the Appendix A or in reference \cite{Goldberg}.
The question we address in this paper is how to estimate a small rotation angle \(\theta_1\) as precisely as possible about a fixed unknown axis \(\boldsymbol{u}\) for which we have no prior knowledge. And, if possible, also gain some knowledge about that axis. In this case, although our system has three parameters, we can treat it as a single-parameter estimation problem. We first consider the problem of estimating \(\theta_1\) when the rotation axis \(\boldsymbol{u}\) is known. 

Let the sensor state be \(\vert\psi\rangle\). It can be shown that the Fisher information\cite{Sidhu} for determining the rotation angle \(\theta_1\) is \cite{Jaspreet},
\begin{equation}\label{Fpsitheta}
    F = 4\left[\langle\psi_{\theta_1}'\vert\psi_{\theta_1}'\rangle-\vert\langle\psi_{\theta_1}'\vert\psi_{\theta_1}\rangle\vert^2\right].
\end{equation}
Here, \(\vert\psi_{\theta_1}'\rangle\) is the derivative of the state \(\vert\psi_{\theta_1}\rangle=\hat{\mathcal{U}}(\theta_1)\vert\psi\rangle\) with respect to parameter \(\theta_1\). By explicitly taking the derivative, one can obtain a simplified form of the Fisher information. Using \(\hat{\mathcal{U}}(\theta_1)=exp(-i\theta_1\boldsymbol{u\cdot\hat{J}})\) and Eq. (\ref{Gk}), we find \(\hat{G}_1\) given by:
\begin{equation}\label{G1}
    \hat{G}_1 = \boldsymbol{u}\cdot\boldsymbol{\hat{J}}.
\end{equation}
Substituting Eq. (\ref{G1}) into Eq. (\ref{Fpsitheta}), we see that the Fisher information for single-parameter estimation depends only on the initial state and the nature of the unitary operation characterized by the generator \(\hat{G}_1\):
\begin{equation}\label{FG1}
    F=4 \left[\langle\psi\vert\hat{G}_1^2\vert\psi\rangle-\langle\psi\vert\hat{G}_1\vert\psi\rangle^2\right].
\end{equation}
An optimization of the Fisher information is then done by adjusting \(\vert\psi\rangle\) such that F is as large as possible. If \(\boldsymbol{u}\) is known, then one can find the eigenstates of \(\hat{G}_1^2\) and choose \(\vert\psi\rangle\) to be an eigenstate of \(\hat{G}_1^2\) with the largest eigenvalue while satisfying \(\langle\hat{G}_1\rangle=0\). 

However, in our case, we do not know \(\boldsymbol{u}\); therefore, we need to consider the effect of the axis \(\boldsymbol{u}\) on determining the Fisher information. To incorporate \(\boldsymbol{u}\) into the existing expression, we insert Eq.(\ref{G1}) into Eq. (\ref{FG1}), and we obtain that F has the explicit form:
\begin{equation}
    F= 4 \left\{\sum_{i,j=1}^3u_i\left[\langle\hat{J}_i\hat{J}_j\rangle-\langle\hat{J}_i\rangle\langle\hat{J}_j\rangle\right]u_j\right\}.
\end{equation}
Define a matrix \(\tilde{\mathcal{J}}\) with the ijth component,
\begin{equation}
    \tilde{\mathcal{J}_{ij}} = \langle\hat{J}_i\hat{J}_j\rangle-\langle\hat{J}_i\rangle\langle\hat{J}_j\rangle.
\end{equation}
The Fisher information is rewritten as a matrix multiplication:
\begin{equation}
    F= 4 \boldsymbol{u}^T\tilde{\mathcal{J}}\boldsymbol{u}.
\end{equation}
Note that \(\tilde{\mathcal{J}}\) is, in general, a Hermitian complex matrix and \(\boldsymbol{u}\) is real; thus we can define a real symmetric matrix:
\begin{equation}\label{Jij}
\begin{split}
    \mathcal{J} &=\frac{\tilde{\mathcal{J}}+\tilde{\mathcal{J}^T}}{2},\\
    \mathcal{J}_{ij}&=\frac{\langle\hat{J}_i\hat{J}_j\rangle+\langle\hat{J}_j\hat{J}_i\rangle}{2}-\langle\hat{J}_i\rangle\langle\hat{J}_j\rangle,
\end{split}
\end{equation}
and replace \(\tilde{\mathcal{J}}\) with \(\mathcal{J}\) for the expression of F without changing the result:
\begin{equation}\label{F as matrix}
    F=4\boldsymbol{u^T}\mathcal{J}\boldsymbol{u}=4\vert\boldsymbol{u}\vert\vert\mathcal{J}\boldsymbol{u}\vert\cos(\omega).
\end{equation}
The parameter \(\omega\) is defined as the unknown angle between \(\boldsymbol{u}\) and \(\mathcal{J}\boldsymbol{u}\). The maximium of the Eq. (\ref{F as matrix}) is obtained when \(\mathcal{J}\boldsymbol{u}\) aligns with \(\boldsymbol{u}\). In other words, the optimal Fisher information is achieved when \(\mathcal{J}\) is diagonal. Now, assume we have no knowledge of \(\boldsymbol{u}\). To keep \(\boldsymbol{u}\) and \(\mathcal{J}\boldsymbol{u}\) always aligned, we would require \(\mathcal{J}\) to have equal diagonal entries. Suppose the sensor state \(\vert\psi\rangle\) has total angular momentum J, by Eq. (\ref{Jij}), we also know that
\begin{equation}
    Tr(\mathcal{J})=\sum_i\langle\hat{J}^2_i\rangle-\langle\hat{J}_i\rangle^2 = J(J+1)-\sum_i\langle\hat{J}_i\rangle^2.
\end{equation}
Hence, the diagonal elements of \(\mathcal{J}\) in the optimal cases is \(\mathcal{J}_{ii}=J(J+1)/3\). Such states satisfy,
\begin{equation}\label{condition anti}
    \begin{split}
        \langle \hat{J}_i^2\rangle&=\frac{J(J+1)}{3},\\
        \langle\hat{J}_i\rangle&=0,
    \end{split}
\end{equation}
are called second-order anti-coherent states or second-order unpolarized states \cite{Goldberg2,Gunnar, Baguette}. The optimal Fisher information is given by:
\begin{equation}
    F = 4 \frac{J(J+1)}{3}.
\end{equation}
A more sophisticated treatment of the multi-parameter Fisher information is given in ref. \cite{Goldberg} and a summary is presented in Appendix A. The method there agrees with the intuitive result presented in this section. The optimal value of the Fisher information given by the Cauchy-Schwarz inequality is referred to as the Quantum Cramer-Rao bound (QCRB).

\section{The Optimal Measurement (For small rotations only)}
We consider the case of a small rotation by \(\theta_1\) around a fixed unknown axis \(\boldsymbol{u}\). Let the initial state be a second-order anti-coherent state \(\vert\phi_0\rangle\). After the rotation, the state is:
\begin{equation}
    \vert \psi_{\boldsymbol{\theta}}\rangle = e^{-i\theta_1(\boldsymbol{\hat{J}\cdot u})}\vert\phi_0\rangle.
\end{equation}
To best estimate the angle \(\theta_1\), we would need to project the rotated state back to the initial state \(\vert\phi_0\rangle\) \cite{Goldberg}. We can define the complete set of orthogonal projectors to be \(\{\hat{\Pi}_\mu\}_{\mu=0}^{2J}\) with \(\hat{\Pi}_0=\vert\phi_0\rangle\langle\phi_0\vert\) and other projectors satisfy \(\hat{\Pi}_\mu\hat{\Pi}_\nu=\delta_{\mu,\nu}\hat{\Pi}_\nu\). To gain information about the rotation axis, note that for small \(\theta_1\) we can perform an expansion for the final state:
\begin{equation}
    \vert\psi_{\boldsymbol{\theta}}\rangle \approx \vert\phi_0\rangle-i\theta_1\boldsymbol{u}\cdot\boldsymbol{\hat{J}}\vert\phi_0\rangle +O(\theta_1)^2.
\end{equation}
The probability of the final state being parallel to \(\hat{J}_i\vert\phi\rangle\) will give us information on \(u_i\). Since the state \(\vert\phi_0\rangle\) is second order anti-coherent satisfy \(\langle\phi_0\vert\hat{J}_i\hat{J}_j\vert\phi_0\rangle=\delta_{ij}J(J+1)/3\), the orthogonal condition is automatically satisfied. In summary, the first four projector operators in the set are:
\begin{equation}
\begin{split}
    \hat{\Pi}_0 &= \vert\phi_0\rangle\langle\phi_0\vert,\\
    \hat{\Pi}_i &= \frac{3}{J(J+1)} \hat{J}_i\vert\phi_0\rangle\langle\phi_0\vert \hat{J}_i=\vert\phi_i\rangle\langle\phi_i\vert.
\end{split}
\end{equation}
The other projection operators (there are J-4 of them) in the set need only be orthogonal to the first four; their expectation values will be small compared to those of the first four, hence can be neglected. We will show that the probability distribution obtained from the first four projection measurement saturates the optimal Fisher information bound for determining \(\theta_1\). 

The Fisher information for a classical multinomial distribution depending on a parameter \(\theta_1\) is given by:
\begin{equation}
    F = \sum_{\mu} \frac{1}{P_{\mu}}\left(\frac{\partial P_\mu}{\partial \theta_1}\right)^2,
\end{equation}
where \(P_\mu\) are the probability of obtaining an outcome labeled by \(\mu\) depending on parameter \(\theta_1\). In our case, these probabilities are generated by:
\begin{equation}\label{Pmu}
P_{\mu}=\langle\psi_{\boldsymbol{\theta}}\vert\hat{\Pi}_\mu\vert\psi_{\boldsymbol{\theta}}\rangle.
\end{equation}
In assuming \(\theta_1\) small, we can expand the probabilities to the second order of \(\theta_1\). We will find the first four probabilities to be:
\begin{equation}\label{Prob approxed}
\begin{split}
    P_0&=1-\theta_1^2\frac{J(J+1)}{3}+O(\theta_1)^3,\\
    P_i&= \theta_1^2u_i^2\frac{J(J+1)}{3}+O(\theta_1)^3.
\end{split}
\end{equation}
Computing the derivatives of Eq. (\ref{Prob approxed}) with respect to \(\theta_1\) we get:
\begin{equation}
    \begin{split}
        \frac{\partial P_0}{\partial\theta_1} &= -2\theta_1\frac{J(J+1)}{3}+O(\theta_1)^2,\\
        \frac{\partial P_i}{\partial\theta_1}&=2\theta_1 u_i^2\frac{J(J+1)}{3}+O(\theta_2)^2.
    \end{split}
\end{equation}
In the asymptotic limit \(\theta_1\rightarrow0\), 
\begin{equation}
\begin{split}
    \lim_{\theta_1\rightarrow0}\frac{1}{P_0}\left(\frac{\partial P_0}{\partial\theta_1}\right)^2 &=0,\\
    \lim_{\theta_1\rightarrow0}\frac{1}{P_i}\left(\frac{\partial P_i}{\partial\theta_1}\right)^2
    &=4u_i^2\frac{J(J+1)}{3}.\\
\end{split}
\end{equation}
The Fisher information for extracting parameter \(\theta_1\) is:
\begin{equation}
    F \approx 4\frac{J(J+1)}{3}\sum_{i=1}^3 u_i^2 = 4\frac{J(J+1)}{3}.
\end{equation}
Thus, we recover the optimal Fisher information we obtained in the previous section. In fact, as long as \(\theta_1\) remains small, using the given set of projection operators, one can recover the optimal Fisher information in the multi-parameter case. Such analysis is given in Appendix B. In the following subsections, we maintain small \(\theta_1\) assumption and drop \(O(\theta_1)^3\) term for the simplicity. The next two subsections present N=4 and N=6 second-order anti-coherent states in the angular-momentum basis that we used in this paper.

 \subsection{N=4 Second Order Anti-coherent State}
A second-order unpolarized state, consisting of four qubits, is a state that satisfies Eq.(\ref{condition anti}) with \(J=2\). An example of a second-order anticoherent state in the angular momentum basis is:
\begin{equation}\label{tetra1}
    \vert \psi_{tetra}\rangle = \frac{1}{\sqrt{3}}(\vert2,2\rangle+\sqrt{2}\vert2,-1\rangle).
\end{equation}
This state is sometimes called a tetrahedron state, as it can be represented as a tetrahedron in the Majorana representation \cite {Goldberg}. However, this state does not correspond to the Bell analysis measurement scheme, in which the Bell qubit pairs allow measurement in an optimal basis. Instead, we consider an initial state, for which the Bell basis analysis gives the approximate result of optimal basis measurement:
\begin{equation}\label{tetra2}
    \vert\psi_0\rangle = \frac{1}{2}(\vert 2,2\rangle+\vert2,-2\rangle+i\sqrt{2}\vert2,0\rangle).
\end{equation}
One can check this by explicitly computing the expectation value in the angular momentum basis given by Eq. (\ref{condition anti}). The measurement basis is:
\begin{widetext}
\begin{equation}
\begin{split}
    &\vert \psi_0\rangle = \frac{1}{2}(\vert 2,2\rangle+\vert2,-2\rangle+i\sqrt{2}\vert2,0\rangle), \\
    &\vert \psi_1\rangle = \frac{1}{\sqrt{2}} \hat{J}_1\vert \psi_0\rangle = e^{i\pi/3}\frac{1}{\sqrt{2}}(\vert2,1\rangle+\vert2,-1\rangle) \equiv \frac{1}{\sqrt{2}}(\vert2,1\rangle+\vert2,-1\rangle),\\
    & \vert \psi_2\rangle =\frac{1}{\sqrt{2}}\hat{J}_2\vert\psi_0\rangle=ie^{-i\pi/3}\frac{1}{\sqrt{2}}(\vert2,1\rangle-\vert2,-1\rangle) \equiv \frac{1}{\sqrt{2}}(\vert2,1\rangle-\vert2,-1\rangle),\\
    & \vert \psi_3\rangle = \frac{1}{\sqrt{2}} \hat{J}_3\vert\psi_0\rangle=\frac{1}{\sqrt{2}}(\vert2,2\rangle-\vert2,-2\rangle).
\end{split}
\end{equation}
\end{widetext}
Bell state analysis can be used to determine the value of the desired set of projection operations.

For now, we would like to show explicitly that our choice of basis recovers the optimal Fisher information \(4J(J+1)/3=8\) for extracting \(\theta_1\). The probabilities are:
\begin{equation}\label{tetra P}
\begin{split}
    & P_0 = 1-2\theta_1^2,\\
    & P_i = 2\theta_1^2u_i^2.\\
\end{split}
\end{equation}
Taking the derivative with respect to \(\theta_1\) gives:
\begin{equation}
\begin{split}
     &\frac{\partial P_0}{\partial\theta_1} = -4 \theta_1,\\
     &\frac{\partial P_i}{\partial\theta_1} = 4 \theta_1 u_i^2.\\
\end{split}
\end{equation}
We can compute the Fisher information for parameter \(\theta_1\):
\begin{equation}\label{tetra fisher}
\begin{split}
    F &\approx \sum_{i=1}^3\frac{1}{P_i}\left(\frac{\partial P_i}{\partial\theta_1}\right)^2=8.
\end{split}
\end{equation}
We see that the basis of our choice has the same enhancement factor of 8 in the case that \(\theta_1\) is small.

\subsection{The N=6 Second Order Anti-coherent State}
Similar to the N=4 case, we make a choice of considering an N=6 second-order anticoherent state for which the Bell basis analysis procedure would work. In the angular momentum notation, the state is given by:
\begin{equation}\label{balance}
    \vert \psi_0\rangle = \frac{1}{\sqrt{2}}(\vert3,2\rangle+\vert3,-2\rangle).
\end{equation}
We can choose the basis for building the projection operators:
\begin{widetext}
\begin{equation}\label{NOON basis}
    \begin{split}
        & \vert \psi_0\rangle = \frac{1}{\sqrt{2}}(\vert3,2\rangle+\vert3,-2\rangle),\\
        & \vert \psi_1\rangle = \frac{1}{2}\hat{J}_1\vert\psi_0\rangle=\frac{1}{4}[\sqrt{3}(\vert3,3\rangle+\vert3,-3\rangle)+\sqrt{5}(\vert3,1\rangle+\vert3,-1\rangle)],\\
        & \vert \psi_2\rangle = \frac{1}{2}\hat{J}_2\vert\psi_0\rangle=\frac{-i}{4}[\sqrt{3}(\vert3,3\rangle-\vert3,-3\rangle)+\sqrt{5}(\vert3,1\rangle-\vert3,-1\rangle)],\\
        & \vert \psi_3\rangle = \frac{1}{2}\hat{J}_3\vert\psi_0\rangle=\frac{1}{\sqrt{2}}(\vert3,2\rangle-\vert3,-2\rangle).\\
    \end{split}
\end{equation}
\end{widetext}
In this case, the probabilities are:
\begin{equation}\label{P6}
\begin{split}
    P_0 &= 1-4\theta_1^2,\\
    P_i &=  4\theta_1^2u_i^2.
    \end{split}
\end{equation}
The corresponding derivatives are:
\begin{equation}
\begin{split}
    \frac{\partial P_0}{\partial \theta_1} &=-8\theta_1,\\
    \frac{\partial P_i}{\partial\theta_1} &=  8\theta_1 u_i^2.\\
\end{split}
\end{equation}
And so the Fisher information for measuring \(\theta_1\) is:
\begin{equation}
    F \approx \sum_{i=1}^3\frac{1}{4\theta_1^2u_i^2}64\theta_1^2u_i^2=16.
\end{equation}
Again, from the formula \(4J(J+1)/3=16\), we see that we recover the enhancement factor of 16 for the N=6 second-order anti-coherent state. The QCRB is saturated in the small \(\theta_1\) regime.

\section{Making Measurements in the Angular Momentum Basis with Linear Optical Components}
\subsection{Bell State Analysis}
The standard Bell state analysis experiment consists of a light source with both the path and polarization degrees of freedom, a 50-50 Beam splitter, two polarizing beam splitters, and photon detectors. To write out the quantum circuit for such an experiment, we need to view the goal of the experiment as distinguishing the antisymmetric Bell state from the symmetric ones. That is to say, the effect of the 50-50 Beam splitter is to perform a polarization-dependent double-controlled gate acting on the path degree of freedom. Let's denote the polarization degree of freedom with orthogonal states \(\vert H\rangle\) and \( \vert V\rangle\); the path degree of freedom with orthogonal states \(\vert u\rangle\) and \(\vert d\rangle\). The Bell states are given by:
\begin{equation}
    \begin{split}
        \vert\phi_0\rangle&=\frac{1}{\sqrt{2}}(\vert HH\rangle+\vert VV\rangle)\otimes\vert ud\rangle,\\
        \vert\phi_1\rangle&=i\frac{1}{\sqrt{2}}(\vert HV\rangle+\vert VH\rangle)\otimes\vert ud\rangle,\\
        \vert\phi_2\rangle&=-\frac{1}{\sqrt{2}}(\vert HV\rangle-\vert VH\rangle)\otimes\vert ud\rangle,\\
        \vert\phi_3\rangle&=i\frac{1}{\sqrt{2}}(\vert HH\rangle-\vert VV\rangle)\otimes\vert ud\rangle.
    \end{split}
\end{equation}
The effect of the 50-50 beam splitter in the Bell state analysis experiment can be represented as the following circuit:
\begin{widetext}

\begin{figure}[h!]
\centering
\begin{quantikz}
\lstick[6]{\(\ket{\phi_\mu}\)}&\qw&\ctrl{1}\gategroup[2,steps=3,style={inner sep=6pt}]{Operation\ 1} &\gate{\hat{H}}   &\qw &\qw &\ctrl{5} \gategroup[6,steps=1,style={inner sep=6pt}]{Operation\ 3} &\qw&\ctrl{1}\gategroup[2,steps=5,style={inner sep=6pt}]{Operation 4}  &\qw      &\gate{\hat{H}} &\qw     &\ctrl{1}&\meter{}\\
&\qw&\targ{}  & \qw  &\qw   &\qw &\ctrl{4} &\qw &\gate{\hat{S}} &\gate{\hat{X}} &\ctrl{-1}&\gate{\hat{X}}&\targ{} &\meter{}\\
\\ \\
&\qw&\gate{\hat{H}}\gategroup[2,steps=2,style={inner sep=6pt}]{Operation\ 2} &\ctrl{1} &\qw &\qw &\gate{\hat{X}}& \qw& \qw& \qw& \qw& \qw& \qw&\meter{}\\
&\qw&\gate{\hat{X}} &\targ{}  &\qw &\qw &\gate{\hat{Z}}& \qw& \qw& \qw& \qw& \qw& \qw& \meter{}
\end{quantikz}
\caption{The quantum circuit represents the idea that a 50-50 beam splitter acts as a polarization-dependent double-control gate acting on the path degree of freedom. Operation 1, 2, and 4 ensures the correct output, and operation 3 captures the polarization-dependent double-control gate.}
\label{fig:placeholder}
\end{figure}

\end{widetext}
Here \(\hat{S}\) represents a phase gate which adds a \(\pi/2\) phase delay to the state \(\vert V\rangle\) to capture the additional phase shift when passing through a beam splitter:
\begin{equation}
    \hat{S}=\begin{pmatrix}
        1&0\\
        0&i
    \end{pmatrix}.
\end{equation}

The top two qubits represent the polarization degree of freedom of the two photons. The bottom two qubits represent the path degree of freedom of the two photons. The ability to distinguish different Bell states comes from making measurements of both the polarization and the path degree of freedom. The setup for Bell-state analysis with linear optical components and identical unitaries applied to each path can be summarized in the following steps. We start with the initial state \(\vert\phi_0\rangle=(\vert HH\rangle+\vert VV\rangle)\otimes\vert ud\rangle/\sqrt{2}\). In the angular momentum basis, the state is \(\vert\phi_0\rangle=(\vert 1,1\rangle+\vert 1,-1\rangle)\otimes\vert ud\rangle/\sqrt{2}\). Here, \(\vert u\rangle\) and \(\vert d\rangle\) represent the two input slots of the 50-50 Beam splitter. Then, we represent the polarization rotation with identical unitaries applied to the two polarization degrees of freedom. In the angular-momentum description, a rotation cannot connect different angular-momentum subspaces labelled by different total angular momenta. This means the state after step one is a linear combination of angular momentum basis states: \(\vert 1,1\rangle, \vert1,0\rangle\), and \(\vert1,-1\rangle\). Alternatively, we can choose the basis which is equivalent to the symmetric Bell states:
\begin{equation}
    \vert\phi_0\rangle=\frac{1}{\sqrt{2}}(\vert1,1\rangle+\vert1,-1\rangle)=\frac{1}{\sqrt{2}}(\vert HH\rangle+\vert VV\rangle),
\end{equation}
\begin{equation}
    \vert\phi_1\rangle=i\vert1,0\rangle = \frac{i}{\sqrt{2}}(\vert HV\rangle+\vert VH\rangle),
\end{equation}
\begin{equation}
    \vert\phi_3\rangle=\frac{i}{\sqrt{2}}(\vert1,1\rangle-\vert1,-1\rangle) = \frac{i}{\sqrt{2}}(\vert HH\rangle-\vert VV\rangle).
\end{equation}
Applying a bit-flip gate, X, on these states results in:

\begin{equation}
\begin{split}
    &\vert\phi_0\rangle\longrightarrow-i\vert\phi_1\rangle=\frac{1}{\sqrt{2}}(\vert VH\rangle+\vert HV\rangle),\\
    &\vert\phi_1\rangle\longrightarrow i\vert\phi_0\rangle= \frac{i}{\sqrt{2}}(\vert VV\rangle+\vert HH\rangle),\\
    &\vert\phi_3\rangle\longrightarrow i\vert\phi_2\rangle = -\frac{i}{\sqrt{2}}(\vert HV\rangle-\vert VH\rangle).
\end{split}
\end{equation}
We note that the standard Bell analysis procedure can fully distinguish these three Bell states, where the 50-50 Beam splitter acts like a polarization-dependent double control gate on the path degree of freedom. By measuring both the polarization and the path taken, we can distinguish the above three states.

\subsection{N=4 Tetrahedron State}
We choose the initial state as the one given by Eq.(\ref{tetra2}), which can be decomposed in the Bell basis as:
\begin{equation}\label{tetrahedron}
    \vert \psi_0 \rangle = \frac{1}{2}[(1+\frac{i}{\sqrt{3}})\vert\phi_0\phi_0\rangle-(1-\frac{i}{\sqrt{3}})\vert \phi_3\phi_3\rangle-\frac{2i}{\sqrt{3}}\vert\phi_1\phi_1\rangle].
\end{equation}
Such a state is generated by the following quantum circuit:
\begin{widetext}

\begin{figure}[h!]
\centering
\begin{quantikz}
&\lstick{$|{0}\rangle$} &\gate{\hat{H}}  &\ctrl{1}  &\qw      &\qw     &\qw&\qw&\qw&\qw&\qw\\
&\lstick{$|{0}\rangle$} &\qw       &\targ{}   &\gate{\hat{Z}} &\qw     &\targ{}&\qw&\qw&\qw&\qw\\
& \lstick{$|{0}\rangle$}&\gate{U_1}&\ctrl{1}  &\ctrl{-1}&\gate{\hat{X}}&\ctrl{-1}&\qw&\qw&\targ{}&\qw\\
&\lstick{$|{0}\rangle$} &\qw       &\gate{U_2}&\ctrl{-2}&\gate{\hat{X}}&\ctrl{-2}&\gate{\hat{X}}&\gate{\hat{H}}&\ctrl{-1}&\qw\\
\end{quantikz}
\caption{The quantum circuit that generates a tetrahedron state given in Eq. (\ref{tetrahedron}).}
\label{fig:placeholder}
\end{figure}

\end{widetext}

The single qubit gates \(U_1\) and \(U_2\) are defined as the following:
\begin{equation}
    U_1=\frac{1}{\sqrt{3}}\begin{pmatrix}
        i&-\sqrt{2}\\
        \sqrt{2}&-i
    \end{pmatrix},
\end{equation}
\begin{equation}
    U_2=\frac{1}{2\sqrt{2}}\begin{pmatrix}
        \sqrt{3}+i&-\sqrt{3}-i\\
        \sqrt{3}-i&\sqrt{3}-i
    \end{pmatrix}.
\end{equation}
Write the corresponding optimal basis states as tensor products of the Bell states:
\begin{equation}\label{tetrabasis}
\begin{split}
    & \vert \psi_0 \rangle = \frac{1}{2}[(1+\frac{i}{\sqrt{3}})\vert\phi_0\phi_0\rangle-(1-\frac{i}{\sqrt{3}})\vert \phi_3\phi_3\rangle-\frac{2i}{\sqrt{3}}\vert\phi_1\phi_1\rangle],\\
     &\vert\psi_1\rangle = \frac{-i}{\sqrt{2}}(\vert\phi_0\phi_1\rangle+\vert\phi_1\phi_0\rangle),\\
    &\vert\psi_2\rangle = \frac{-1}{\sqrt{2}}(\vert\phi_3\phi_1\rangle+\vert\phi_1\phi_3\rangle),\\
    &\vert\psi_3\rangle = \frac{-i}{\sqrt{2}}(\vert\phi_0\phi_3\rangle+\vert\phi_3\phi_0\rangle),\\
    & \vert\psi_4\rangle = \frac{1}{2}[(1-\frac{i}{\sqrt{3}})\vert\phi_0\phi_0\rangle-(1+\frac{i}{\sqrt{3}})\vert \phi_3\phi_3\rangle+\frac{2i}{\sqrt{3}}\vert\phi_1\phi_1\rangle].
\end{split}
\end{equation}

Suppose the final state is given by:
\begin{equation}
    \vert \psi\rangle=\sum_{\mu=0}^4 \alpha_\mu\vert\psi_\mu\rangle.
\end{equation}
Let's denote the projection to state \(\vert\phi_i\phi_j\rangle\) as:
\begin{equation}
    P_{ij} = \vert\langle\psi\vert\phi_i\phi_j\rangle\vert^2.
\end{equation}
From Eq. (\ref{tetrabasis}) we find:
\begin{equation}
    P_0\approx P_0+P_4 = P_{00}+P_{33}+P_{11},
\end{equation}
\begin{equation}
\begin{split}
        & P_1 = P_{01}+P_{10},\\
        & P_2 = P_{13}+P_{31},\\
        & P_3 = P_{03}+P_{30}.
\end{split}
\end{equation}
We made the approximation \(P_0\approx P_0+P_4\) because according to Eq. (\ref{tetra P}), \(P_0+\sum_{i=1}^3P_i=1\) in the \(O(\theta_1)^2\) order of approximation, meaning that \(P_4\) has at least other \(O(\theta_1)^3\) which can be ignored. This means that even though we cannot make direct projection measurements in the optimal basis, projection measurements on pairs of Bell states are sufficient to extract the required probabilities.

\subsection{The N=6 Second Order Anti-coherent State}
We can perform a similar decomposition for an N=6 second-order anti-coherent state, referred to as the Balanced N=6 NOON state, given by Eq. (\ref{balance}). We can write the complete angular momentum basis state in terms of the tensors of Bell states:
\begin{widetext}
\begin{equation}\label{6qubit}
\begin{split}
    \vert\psi_0\rangle &= \frac{-i}{\sqrt{6}}(\vert\phi_0\phi_0\phi_1\rangle-\vert\phi_3\phi_3\phi_1\rangle
    +\vert\phi_1\phi_0\phi_0\rangle-\vert\phi_1\phi_3\phi_3\rangle
    +\vert\phi_0\phi_1\phi_0\rangle-\vert\phi_3\phi_1\phi_3\rangle),\\
    \vert\psi_1\rangle&=\frac{1}{2\sqrt{2}}[\sqrt{3}\vert\phi_0\phi_0\phi_0\rangle
    -\frac{1}{\sqrt{3}}(\vert\phi_3\phi_3\phi_0\rangle+\vert\phi_0\phi_3\phi_3\rangle+\vert\phi_3\phi_0\phi_3\rangle)\\
    &-\frac{2}{\sqrt{3}}(\vert\phi_0\phi_1\phi_1\rangle+\vert\phi_1\phi_0\phi_1\rangle+\vert\phi_1\phi_1\phi_0\rangle)],\\
    \vert\psi_2\rangle &=\frac{1}{\sqrt{6}}[\vert\phi_0\phi_3\phi_0\rangle+\vert\phi_3\phi_0\phi_0\rangle+\vert\phi_0\phi_0\phi_3\rangle-\vert\phi_3\phi_1\phi_1\rangle-\vert\phi_1\phi_3\phi_1\rangle-\vert\phi_1\phi_1\phi_3\rangle],\\
    \vert\psi_3\rangle &= \frac{-1}{\sqrt{6}}(\vert\phi_0\phi_3\phi_1\rangle+\vert\phi_3\phi_0\phi_1\rangle
    +\vert\phi_1\phi_0\phi_3\rangle+\vert\phi_1\phi_3\phi_0\rangle +\vert\phi_0\phi_1\phi_3\rangle+\vert\phi_3\phi_1\phi_0\rangle),\\
    \vert\psi_4\rangle& = \frac{-i}{\sqrt{10}}(\vert\phi_0\phi_0\phi_1\rangle-\vert\phi_3\phi_3\phi_1\rangle
    +\vert\phi_1\phi_0\phi_0\rangle-\vert\phi_1\phi_3\phi_3\rangle
    +\vert\phi_0\phi_1\phi_0\rangle-\vert\phi_3\phi_1\phi_3\rangle+2\vert\phi_1\phi_1\phi_1\rangle),\\ 
    \vert\psi_5\rangle&=\frac{1}{2\sqrt{2}}[\sqrt{\frac{1}{5}}\vert\phi_0\phi_0\phi_0\rangle
    -\frac{3}{\sqrt{5}}(\vert\phi_3\phi_3\phi_0\rangle+\vert\phi_0\phi_3\phi_3\rangle+\vert\phi_3\phi_0\phi_3\rangle)\\
    &+\frac{2}{\sqrt{5}}(\vert\phi_0\phi_1\phi_1\rangle+\vert\phi_1\phi_0\phi_1\rangle+\vert\phi_1\phi_1\phi_0\rangle)],\\
    \vert\psi_6\rangle &=\frac{1}{\sqrt{2}}[\frac{1}{\sqrt{5}}(\vert\phi_0\phi_3\phi_0\rangle+\vert\phi_3\phi_0\phi_0\rangle+\vert\phi_0\phi_0\phi_3\rangle\\
    &-\vert\phi_3\phi_1\phi_1\rangle-\vert\phi_1\phi_3\phi_1\rangle-\vert\phi_1\phi_1\phi_3\rangle)+\frac{2}{\sqrt{5}}\vert\phi_3\phi_3\phi_3\rangle].\\
\end{split}
\end{equation}

\noindent The Quantum circuit to generate such a state might be written as in Figure 3.

\begin{figure}[h!]

\centering
\begin{adjustbox}{width=\columnwidth}

\begin{quantikz}
&\lstick{$|{0}\rangle$}&\gate{\hat{H}}&\ctrl{1}&\qw&\qw&\qw&\qw&\qw&\qw&\qw&\qw&\qw&\qw&\qw&\qw&\qw&\qw&\qw&\qw&\qw&\qw&\qw\\
&\lstick{$|{0}\rangle$}&\qw&\targ{} &\targ{}&\qw&\qw&\qw&\qw&\gate{\hat{Z}}&\qw&\qw&\qw&\qw&\targ{}&\qw&\gate{\hat{Z}}&\qw&\qw&\qw&\qw&\qw&\qw\\
&\lstick{$|{0}\rangle$}&\gate{\hat{H}}&\ctrl{1}&\ctrl{-1}&\ctrl{1} &\targ{}&\targ{}&\qw&\ctrl{-1}&\ctrl{1}&\gate{\hat{H}}&\ctrl{1}&\qw&\ctrl{-1}&\targ{}&\ctrl{-1}&\qw&\targ{}&\qw&\qw&\qw&\qw\\
&\lstick{$|{0}\rangle$}&\qw&\targ{}&\qw&\targ{}&\gate{\hat{H}}&\ctrl{-1}&\qw&\qw&\targ{}&\qw&\targ{}&\qw&\qw&\qw&\qw&\gate{\hat{H}}&\ctrl{-1}&\qw&\qw&\qw&\qw\\
&\lstick{$|{0}\rangle$}&\gate{\hat{U}}&\ctrl{1}&\ctrl{-2}&\ctrl{-2}&\ctrl{-2}&\ctrl{-1}&\qw&\ctrl{-2}&\ctrl{-1}&\ctrl{-2}&\ctrl{-1}&\gate{\hat{X}}&\ctrl{-2}&\ctrl{-2}&\ctrl{-2}&\ctrl{-1}&\ctrl{-1}&\ctrl{1}&\gate{\hat{H}}&\ctrl{1}&\qw\\
&\lstick{$|{0}\rangle$}&\qw&\gate{\hat{H}}&\ctrl{-1} &\ctrl{-1}&\ctrl{-1}&\ctrl{-1}&\gate{\hat{X}}&\ctrl{-1}&\ctrl{-1}&\ctrl{-1}&\ctrl{-1}&\qw&\ctrl{-1}&\ctrl{-1}&\ctrl{-1}&\ctrl{-1}&\ctrl{-1}&\targ{}&\qw&\targ{}&\qw\\
\end{quantikz}
\end{adjustbox}
\caption{The quantum circuit that generates a six-qubit second-order anti-coherent state given by \(\vert\psi_0\rangle\) in Eq. (\ref{6qubit}).}
\label{fig:placeholder}
\end{figure}

\end{widetext}
Here, the gate U is given by the matrix:
\begin{equation}
    U=\frac{1}{\sqrt{3}}\begin{pmatrix}
      1&-\sqrt{2}\\
      \sqrt{2}&1
    \end{pmatrix}.
\end{equation}
Then, use the result of the Bell state analysis to extract the expectation value of the required projection operator. For example, suppose the final state is:
\begin{equation}
    \vert\psi\rangle=\sum_{\mu=0}^6\alpha_\mu\vert\psi_\mu\rangle.
\end{equation}
Define the projection on to the state \(\vert\phi_i\phi_j\phi_p\rangle\) as:
\begin{equation}
    P_{ijk}=\vert\langle\phi_i\phi_j\phi_p\vert\psi\rangle\vert^2.
\end{equation}
We find:
\begin{equation}
    P_{100}=P_{010}=P_{001}=\vert\frac{1}{\sqrt{6}}\alpha_0+\frac{1}{\sqrt{10}}\alpha_4\vert^2,
\end{equation}
\begin{equation}
    P_{133}=P_{313}=P_{331}=\vert\frac{1}{\sqrt{6}}\alpha_0-\frac{1}{\sqrt{10}}\alpha_4\vert^2,
\end{equation}
\begin{equation}
    P_{111}=\frac{2}{5}\vert
    \alpha_4\vert^2.
\end{equation}
Using the same approximation from the previous section, when kept to the 2nd order of \(\theta_1\):
\begin{equation}
    1 = P_0+P_1+P_2+P_3.
\end{equation}
Therefore, \(P_4,P_5,P_6 \approx0\). An approximate value of \(P_0\) is:
\begin{widetext}
\begin{equation} 
P_0 \approx P_0+P_4=P_{100}+P_{010}+P_{001}+P_{133}+P_{313}+P_{331}+P_{111}.
\end{equation}
where \(P_4\) is assumed to be neglected since it has at least an order of \(O(\theta_1)^3\). Similarly,
\begin{equation}
    \begin{split}
        & P_1 \approx P_1+P_5 = P_{000}+P_{330}+P_{033}+P_{303}+P_{011}+P_{101}+P_{110},\\
        & P_2 \approx P_2+P_6 = P_{030}+P_{300}+P_{003}+P_{311}+P_{131}+P_{311},\\
        & P_3 = P_{031}+P_{301}+P_{103}+P_{130}+P_{013}+P_{310}.
    \end{split}
\end{equation}
\end{widetext}

\section{Extracting the Parameters and Computing the Variances}
The parameters \(\boldsymbol{u}\) and \(\theta_1\) can be extracted from Eq. (\ref{tetra P}) and Eq. (\ref{P6}) accordingly:
\begin{equation}
\begin{split}
     &\vert\theta_1\vert = \sqrt{\frac{1-P_0}{2}},\\
     &\vert u_i\vert = \sqrt{\frac{P_i}{1-P_0}}.
\end{split}
\end{equation}
Note that the procedure for multi-parameter estimation suffers from the same problem as the single-phase estimation scheme, where the parameters are decided up to a \(\pm\)
sign. This means that the procedures provided are more relevant in estimating the magnitude of \(\theta_1\) around an unknown axis than in providing full information regarding the polarization rotation.

The count statistic of the procedure we provided in the previous sections follows a multinomial distribution. We denote the possible outcomes as \(\chi_i\) corresponding to  \(\vert\psi_i\rangle\), with probability \(P_i\). When n independent experiments are performed, it has the following properties \cite{Shanmugam}:
\begin{widetext}
\begin{equation}\label{multi}
\begin{split}
     &{\rm Var}(\chi_i)=nP_i(1-P_i),\\
     &{\rm Cov}(\chi_i,\chi_j)=-nP_iP_j,\\
     &{\rm Var}\left(\sum_i a_i\chi_i\right)=\sum_i a_i^2{\rm Var}(\chi_i)+2\sum_i\sum_{j>i}a_ia_j{\rm Cov}(\chi_i,\chi_j),\\
     &{\rm Var}\left(\sum_i\chi_i\right) = n\left(\sum_iP_i\right)\left(1-\sum_iP_i\right).\\
\end{split}
\end{equation}
\end{widetext}
This allows us to compare the variance of the Bell basis with that of the optimal basis. 

For the N=4 tetrahedron state, the variance for \(\chi_0\) is:
\begin{equation}
    {\rm Var}(\chi_0) = nP_0(1-P_0).
\end{equation}
Expand the variance in \(\theta_1\) and kept to \(O(\theta_1)^2\):
\begin{equation}
    {\rm Var}(\chi_0) \approx n(1-2\theta_1^2)(1-1+2\theta_1^2)\approx 2n\theta_1^2.
\end{equation}
We would like to compare the variance above to the one resulting from measuring in the Bell basis. Let's denote the Bell basis outcome as \(\tilde{\chi}_{ij}\) where i,j correspond to the Bell state subscripts. For example, \(\tilde{\chi}_{00}\) denote the out come of obtaining \(\vert\phi_0\phi_0\rangle\). Then \(\chi_0+\chi_4\) can be written as:
\begin{equation}
    \chi_0+\chi_4 = \tilde{\chi}_{00}+\tilde{\chi}_{33}+\tilde{\chi}_{11}.
\end{equation}
The variance calculated following Eq. (\ref{multi}) is:
\begin{equation}
\begin{split}
    &{\rm Var}(\chi_0+\chi_4) \\&= n(P_{00}+P_{33}+P_{11})(1-P_{00}-P_{33}-P_{11})\\
    &= n(P_0+P_4)(1-P_0-P_4).
\end{split}
\end{equation}
Let's expand the variance in \(\theta_1\) and kept to \(O(\theta_1)^2\):
\begin{equation}
    \begin{split}
    &{\rm Var}(\chi_0+\chi_4) \\&\approx n[(1-2\theta_1^2)+O(\theta_1)^3](1-1+2\theta_1^2-O(\theta_1)^3)\\
    & \approx 2n\theta_1^2.
\end{split}
\end{equation}
The error resulting from using the Bell basis is negligible for higher orders in \(\theta_1\). However, if one keeps higher-order terms in the expansion, the error is expected to increase. 

The enhancement in measuring the parameter \(\theta_1\) can be calculated directly from the error-propagation formula.
\begin{equation}
    \sigma_{\theta_1} = \left \vert \frac{\partial \theta_1}{\partial P_0} \right\vert \sigma_{P_0} \approx \frac{1}{2\sqrt{2}}\frac{1}{\sqrt{1-1+2\theta_1^2}}\frac{\sqrt{2\theta_1^2}}{\sqrt{n}}=\frac{1}{2\sqrt{2n}}.
\end{equation}
Where we keep to order \(O(\theta_1)^2\) and n is the number of independent trials. Recall the Fisher information in the optimal basis is given by Eq. (\ref{tetra fisher}) where \(F\approx8\). For n independent experiments, the standard deviation is:
\begin{equation}
    \sigma_{\theta_1} = \frac{1}{\sqrt{nF}} = \frac{1}{2\sqrt{2n}}.
\end{equation}
Thus, we have shown that under the approximation for small \(\theta_1\), the Bell basis has the same variance as the optimal basis and saturates the QCRB for estimating \(\theta_1\).

\section{Conclusion}
In this paper, we present a procedure for making optimal measurements regarding small rotations around an unknown axis that saturate the Quantum Cramér–Rao Bound (QCRB) for N = 4 and N = 6 anti-coherence states given by:
\begin{equation}
    \vert \psi_{tetra}\rangle = \frac{1}{2}(\vert2,2\rangle+\vert2,-2\rangle+i\sqrt{2}\vert2,0\rangle),
\end{equation}
\begin{equation}
    \vert\psi_{balance}\rangle = \frac{1}{\sqrt{2}}(\vert3,2\rangle+\vert3,-2\rangle).
\end{equation}
Our work suggests that achieving an optimal measurement is possible. Assuming the initial state can be generated using solid-state sources such as neural atom arrays or quantum dots, the parameter extraction is not experimentally demanding; our scheme requires only linear optical components. We demonstrate, through explicit computation, that the QCRB is saturated using the multinomial distribution. The standard deviation for rotation \(\theta_1\) around an unknown axis is:
\begin{equation}
\sigma_{\theta_1} = \frac{1}{\sqrt{\frac{4J(J+1)}{3}n}}.
\end{equation}
for \(J=2\) and \(J=3\) where n is the number of trials. We presented the quantum circuit for generating the second-order anti-coherent states. The quantum circuit requires many multi-qubit gates even for \(N=4\) and \(N=6\). The model suffers from scaling issues. The procedure requires the amount of rotation featured by \(\theta_1\) to be small; any term in the expansion with \(O(\theta_1)^2\) has to be negligible. To find the acceptable range of rotation, one would need to calculate the exact expectation values and compare them with the approximation. Such an approach requires extensive numerical simulation and is beyond the scope of our paper's analysis. Additionally, each photon in the anti-coherence state must follow a separate path, which increases the difficulty of the state-generation step. Like all quantum-enhanced metrology results, the procedure we presented is very sensitive to the initial state condition and noise. Since we assume an unknown rotation axis, our model will not be able to satisfy the
“Hamiltonian-not-in-Kraus-span” (HNKS) condition \cite{zhou}. Hence, in the presence of a significant noise channel, or a non-ideal initial state, we expect no quantum enhancement. We do not claim that our procedure is the most resource-efficient; however, in the absence of research in this area, our work may suggest the types of states required for attaining the QCRB. Finally, as an outlook, we suggest it may be possible to incorporate adaptive tomography techniques into our approach to overcome the limitation of small rotations. We first estimate the parameters using conventional methods, then use second-order anti-coherent states with optimal basis measurements to improve accuracy.

\appendix
\begin{widetext}
\section{Multi-parameter Fisher Information}
Even in the general case, the Fisher information can still be written in expectation values of the generators \(\hat{G}_k\) corresponding to the parameters \(\theta_k\). It can be shown that \(\hat{G}_k\) takes on the form\cite{Goldberg, Wilcox, Suzuki, Hou}:
\begin{equation}
    \hat{G}_k = \boldsymbol{g}_k(\boldsymbol{\theta}) \cdot \boldsymbol{\hat{J}},
\end{equation}
where \(\boldsymbol{\hat{J}}\) is the angular momentum operator written as a vector and \(\boldsymbol{g}_k(\boldsymbol{\theta})\) is a set of coefficients which has an expression that depends on the choice of the rotation parameters.
\begin{equation}\label{g(theta)k}
\begin{split}
    & \boldsymbol{g_1}(\boldsymbol{\theta}) = \boldsymbol{u},\\
    & \boldsymbol{g_2}(\boldsymbol{\theta}) = \sin{(\theta_1)}[\cos{(\theta_1)}\frac{\partial\boldsymbol{u}}{\partial\theta_2}-\sin{(\theta_1)\frac{\partial\boldsymbol{u}}{\partial\theta_2}}\times\boldsymbol{u}],\\
    & \boldsymbol{g_3}(\boldsymbol{\theta}) = \sin{(\theta_1)}[\cos{(\theta_1)}\frac{\partial\boldsymbol{u}}{\partial\theta_3}-\sin{(\theta_1)\frac{\partial\boldsymbol{u}}{\partial\theta_3}}\times\boldsymbol{u}].\\
\end{split}
\end{equation}
The parameterization provided above is just one example of the many choices available. We will provide a derivation of \(\boldsymbol{g(\theta)}\) below. Recall that the unitary operator for polarization rotation \(\hat{\mathcal{U}}\) can be written as a tensor of a single-qubit unitary:
\begin{equation}\label{U=U}
    \hat{\mathcal{U}}= e^{-i\theta\boldsymbol{u\cdot\hat{J}}} = \hat{U}
^{\otimes n}=\left[e^{-i\theta\boldsymbol{u\cdot\hat{\sigma}}}\right]^{\otimes n}.
\end{equation}
Therefore, by definition \(\hat{G}_k\) given in Eq. (\ref{Gk}) and use Eq. (\ref{U=U}), generators \(\hat{G}_k\) are:
\begin{equation}
    \hat{G}_k = i \sum_{per}\left[\frac{d\hat{U}}{d\theta_k}\hat{U}^{\dagger}\right]^{(l)}\otimes\hat{I}^{\otimes (n-1)}.
\end{equation}
As in the main text, we use a superscript l to represent a qubit operator that is applied to a qubit at position l, and the sum is over all possible permutations of qubit position. Now, we can use the property of a unitary operation on a qubit to compute the operator at each position l. Notice that for unitary rotation \(\hat{U}\):
\begin{equation}
    \hat{U}=e^{-i\theta_1\boldsymbol{u}(\theta_2,\theta_3)\boldsymbol{\cdot\hat{\sigma}}}=\cos(\theta_1)\hat{I}-i\sin(\theta_1)\boldsymbol{u\cdot\sigma}.
\end{equation}
Hence, we have the expression:
\begin{equation}
    \frac{d\hat{U}}{d\theta_k}\hat{U}^{\dagger} = \left\{\frac{d}{d\theta_k}\left[\cos(\theta_1)\hat{I}-i\sin(\theta_1)\boldsymbol{u\cdot\hat{\sigma}}\right]\right\}\left[\cos(\theta_1)\hat{I}+i\sin(\theta_1)\boldsymbol{u\cdot\hat{\sigma}}\right].
\end{equation}
For \(\theta_1\), we have:
\begin{equation}
    \frac{d\hat{U}}{d\theta_1}\hat{U}^{\dagger} =-i\boldsymbol{u\cdot\hat{\sigma}}.
\end{equation}
For \(k=2,3\), we have:
\begin{equation}\label{UG2}
    \frac{d\hat{U}}{d\theta_k}\hat{U}^{\dagger} = \left[-i\sin(\theta_1)\frac{d\boldsymbol{u}}{d\theta_k}\boldsymbol{\cdot\hat{\sigma}}\right]\left[\cos(\theta_1)\hat{I}+i\sin(\theta_1)\boldsymbol{u\cdot\hat{\sigma}}\right] = -i\sin(\theta_1)\cos(\theta_1)\left(\frac{d\boldsymbol{u}}{d\theta_k}\boldsymbol{\cdot\hat{\sigma}}\right)+\sin^2(\theta)\left(\frac{d\boldsymbol{u}}{d\theta_k}\boldsymbol{\cdot u}\right)\left(\boldsymbol{u\cdot\hat{\sigma}}\right).
\end{equation}
Using the property for the Pauli operator:
\begin{equation}
    (\boldsymbol{a\cdot\hat{\sigma}})(\boldsymbol{b\cdot\hat{\sigma}})=(\boldsymbol{a\cdot b})\hat{I}+i(\boldsymbol{a\times b})\cdot\boldsymbol{\hat{\sigma}}.
\end{equation}
And notice that \(\boldsymbol{u}\) given in Eq. (\ref{u}) is a unit vector, hence pointing in the radial direction of a unit sphere. Therefore, varying the direction of \(\boldsymbol{u}\) by changing \(\theta_k\) would result in a change in the orthogonal direction to \(\boldsymbol{u}\). This means,
\begin{equation}
    \frac{d\boldsymbol{u}}{d\theta_k}\cdot\boldsymbol{u} = 0.
\end{equation}
We can simplify Eq. (\ref{UG2}), so we obtained:
\begin{equation}
    \frac{d\hat{U}}{d\theta_k}\hat{U}^{\dagger} = -i\left[\sin(\theta_1)\cos(\theta_1)\frac{d\boldsymbol{u}}{d\theta_k} -\sin^2(\theta_1)\left(\frac{d\boldsymbol{u}}{d\theta_k}\times\boldsymbol{u}\right)\right]\cdot\boldsymbol{\hat{\sigma}}.
\end{equation}
Substituting these back into the angular momentum representation, we derived the expression for Eq. (\ref{g(theta)k}).

The Fisher information matrix for multi-parameter estimation is formulated in the covariance of generators \(\hat{G}_k\). Let the density matrix of the quantum state after rotation be \(\hat{\rho}_{\boldsymbol{\theta}} \). The Fisher information matrix is given by \cite{Goldberg}:
\begin{equation}\label{fisher1}
    Q_{ij}(\hat{\rho}_{\boldsymbol{\theta}}) = 4 Cov_{\hat{\rho}_{\boldsymbol{\theta}}} (\hat{G}_i, \hat{G}_j),
\end{equation}
where \(Cov_{\hat{\rho}_{\boldsymbol{\theta}}}\) refer to taking the covariance of generators \(\hat{G}_i\) and \(\hat{G}_j\) with respect to the final state:
\begin{equation}\label{cov}
    Cov_{\hat{\rho}_{\boldsymbol{\theta}}}(\hat{G}_i,\hat{G}_j) = Tr\{\hat{\rho}_{\boldsymbol{\theta}}[(\hat{G}_i-Tr(\hat{\rho}_{\boldsymbol{\theta}}\hat{G}_i))(\hat{G}_j-Tr(\hat{\rho}_{\boldsymbol{\theta}}\hat{G}_j))
    +(\hat{G}_j-Tr(\hat{\rho}_{\boldsymbol{\theta}}\hat{G}_j))(\hat{G}_i-Tr(\hat{\rho}_{\boldsymbol{\theta}}\hat{G}_i))]\}.
\end{equation}

Since we have a pure state that undergoes a unitary process, we can view the polarization rotation as a three-by-three real rotation matrix acting on the angular-momentum operators. Let the rotation matrix be \(R(\boldsymbol{\theta})\), the angular momentum operators transform as vectors:
\begin{equation}
    \hat{\mathcal{U}}^\dagger \boldsymbol{\hat{J}} \hat{\mathcal{U}} = R(\boldsymbol{\theta})\boldsymbol{\hat{J}}.
\end{equation}
Define a matrix \(\mathfrak{G}(\boldsymbol{\theta})\) with matrix elements given by \(\mathfrak{G}_{ij}=\left[\boldsymbol{g}_i(\boldsymbol{\theta})\right]_j \). Let the density matrix of the initial pure state be \(\hat{\rho}=\vert\psi\rangle\langle\psi\vert\), we can define a three-by-three sensitivity matrix which is the covariance matrix of angular momentum operators \(\hat{J}_m, \hat{J}_n\) with respect to the initial state:
\begin{equation}
    C_{ij}(\hat{\rho}) = \frac{1}{2}\langle\hat{J}_{i}\hat{J}_j+\hat{J}_{j}\hat{J}_i\rangle -\langle \hat{J}_i\rangle\langle\hat{J}_j\rangle.
\end{equation}
By explicit computation following Eq. (\ref{cov}) and (\ref{fisher1}), one can show that the quantum Fisher information matrix is\cite{Goldberg}:
\begin{equation}
    Q(\hat{\rho}_{\boldsymbol{\theta}}) = 4\mathfrak{G}(\boldsymbol{\theta})^T R(\boldsymbol{\theta})^TC(\hat{\rho})R(\boldsymbol{\theta})\mathfrak{G}(\boldsymbol{\theta})
    = 4\tilde{\mathfrak{G}}(\boldsymbol{\theta})^TC(\hat{\rho})\tilde{\mathfrak{G}}(\boldsymbol{\theta}).
\end{equation}

The Fisher information is maximized for
second-order unpolarized states which satisfy:
\begin{equation}\label{anti}
\begin{split}
\langle\hat{J_i}\hat{J_j}\rangle&=\delta_{ij}\frac{J(J+1)}{3},\\
\langle \hat{J}_i\rangle&=0.
\end{split}
\end{equation}
These states are maximally entangled and are sometimes referred to as the 'Kings of quantumness' states \cite{Bouchard}.
For such states, we obtain a simple expression for the sensitivity matrix:
\begin{equation}
    C_{ij}(\hat{\rho}_{anti}) = \frac{J(J+1)}{3}\delta_{ij}.
\end{equation}
Where I is the identity matrix. Also, \(R(\boldsymbol{\theta})\) is a 3D rotation matrix, so \(R(\boldsymbol{\theta})^TR(\boldsymbol{\theta})=I\).
Therefore,
\begin{equation}\label{Q pure}
Q(\hat{\rho}_{\boldsymbol{\theta}}) = 4\left[\frac{1}{3}J(J+1)\right]\mathfrak{G}(\boldsymbol{\theta})^T R(\boldsymbol{\theta})^TR(\boldsymbol{\theta})\mathfrak{G}(\boldsymbol{\theta}) = \frac{4J(J+1)}{3}\mathfrak{G}(\boldsymbol{\theta})^T\mathfrak{G}(\boldsymbol{\theta}).
\end{equation}

\section{The Optimal Projection Basis in the Multi-parameter Case}
The expectation values of the projector operators are the same as those defined in Eq. (\ref{Pmu}). However, we would use the vectors \(\boldsymbol{g}_k(\boldsymbol{\theta})\) given in Appendix A to express the derivatives of these expectation values:
\begin{equation}
    \frac{\partial P_{\mu}}{\partial\theta_k} = \langle\psi_{\boldsymbol{\theta}}\vert\frac{\partial U^{\dagger}}{\partial\theta_k}\hat{\Pi}_{\mu}\vert\psi_{\boldsymbol{\theta}}\rangle + c.c.
     = i\boldsymbol{g}_k(\boldsymbol{\theta})\cdot\langle\psi_{\boldsymbol{\theta}}\vert \boldsymbol{\hat{J}}\hat{\Pi}_{\mu}\vert\psi_{\boldsymbol{\theta}}\rangle + c.c..
\end{equation}
With the small rotation angle approximation, we can write the final state as the expression below:
\begin{equation}
    \vert \psi_{\boldsymbol{\theta}}\rangle \approx (1-i\theta_1(\boldsymbol{\hat{J}}\boldsymbol{\cdot u})-\frac{1}{2}\theta_1(\boldsymbol{\hat{J}}\cdot\boldsymbol{u})^2)\vert\phi_0\rangle.
\end{equation}
Recall that the initial state \(\vert\phi_0\rangle\) is a second-order unpolarized state, which means we can use Eq. (\ref{anti}) to further simplify our computation. Keep to the second order of \(\theta_1\) when computing the expectation value, we find the probabilities:
\begin{equation}\label{pk}
\begin{split}
    P_0&= 1-\theta_1^2\langle\phi_0\vert(\boldsymbol{\hat{J}}\cdot\boldsymbol{u})^2\vert\phi_0\rangle =1-\frac{1}{3}J(J+1)\theta_1^2,\\
    P_i &= \frac{1}{\langle\phi_0\vert \hat{J}_i^2\vert\phi_0\rangle}\theta_1^2\vert\langle\phi_0\vert(\boldsymbol{\hat{J}}\cdot\boldsymbol{u})\hat{J}_i\vert\phi_0\rangle\vert^2
    = \frac{1}{3}J(J+1)\theta_1^2u_i^2.\\
\end{split} 
\end{equation}
The corresponding derivatives, kept to the 1st order of \(\theta_1\) are:
\begin{equation}\label{dp0}
    \frac{\partial P_0}{\partial\theta_k} = 2\theta_1\langle\phi_0\vert(\boldsymbol{u}\cdot\boldsymbol{\hat{J}})(\boldsymbol{g}_k(\boldsymbol{\theta})\cdot\boldsymbol{\hat{J}})\vert\phi_0\rangle
    =2\theta_1\frac{1}{3}J(J+1)(\boldsymbol{u}\cdot\boldsymbol{g}_k(\boldsymbol{\theta})),
\end{equation}
\begin{equation}\label{dpl}
    \frac{\partial P_i}{\partial\theta_k} = 2\theta_1\langle\phi_0\vert(\boldsymbol{g}_k(\boldsymbol{\theta})\cdot\boldsymbol{\hat{J}})\hat{\Pi}_{\mu}(\boldsymbol{u}\cdot\boldsymbol{\hat{J}})\vert\phi_0\rangle\\
    =2\theta_1\frac{1}{3}J(J+1)u_i(\boldsymbol{g}_k(\boldsymbol{\theta}))_i.
\end{equation}
We noticed that the probabilities \(P_{\mu}\) are functions of \(\theta_1\) and \(\boldsymbol{u}\). Therefore, we can extract the parameter of rotation from the probabilities \(P_{\mu}\). We also noticed that:
\begin{equation}
     \frac{1}{P_0}\left(\frac{\partial P_0}{\partial\theta_k}\right)^2
     = \frac{4J^2(J+1)^2(\boldsymbol{u\cdot g_k}(\theta_1))^2}{9}\frac{\theta_1^2}{\left(1-\frac{J(J+1)\theta_1^2}{3}\right)},
\end{equation}
\begin{equation}
    \sum_{i=1}^3\frac{1}{P_i}\left(\frac{\partial P_i}{\partial\theta_k}\right)^2= \frac{4J(J+1)}{3}\sum_{i=1}^3(\boldsymbol{g}_k(\boldsymbol{\theta}))_i^2.
\end{equation}
And so the contribution from \(P_0\) term in the Fisher information is \(\theta_1^2\) order smaller than the \(P_l\) terms. We can approximate the Fisher information for parameter \(\theta_k\) as:
\begin{equation}
\begin{split}
    F(\theta_k) = \sum_{\mu=0}^3 \frac{1}{P_\mu}\left(\frac{\partial P_\mu}{\partial\theta_k}\right)^2 \approx\sum_{i=1}^3\frac{1}{P_i}\left(\frac{\partial P_i}{\partial\theta_k}\right)^2
    \approx 4\frac{J(J+1)}{3}\sum_{i=1}^3(g_k(\boldsymbol{\theta}))_i^2
    = 4 \mathfrak{G}(\boldsymbol{\theta})^TC(\hat{\rho})\mathfrak{G}(\boldsymbol{\theta})= Q(\hat{\rho})_{kk}.
\end{split}
\end{equation}
Thus, we have shown that our choice of basis saturates the Quantum Cramér–Rao bounds when the rotation is small.
\end{widetext}
\end{document}